\documentstyle[12pt, epsf]{article}
\begin{document}

\baselineskip=21.5pt

\begin{flushleft}

{\Large {\bf 
A Mathematical Model with Modified Logistic  
}}
{\Large {\bf 
Approach for Singly-Peaked Population Processes 
}}

\vskip 1.0cm

{\large Ryoitiro Huzimura$^{*1}$ and Toyoki Matsuyama$^{\dagger2}$}

\begin{it}
$^*$Department of Economics, Osaka Gakuin University, 
2-36-1 Kishibe-minami, Suita-shi, Osaka 564-8511, Japan 
and 
$^\dagger$Department of Physics, Nara University of Education, 
Takabatake-cho, Nara 630-8528, Japan 
\end{it}

\vskip 5.0cm

$^1$Fax: 0081-06-6382-4363. 

$^2$Fax: 0081-0742-27-9289.  E-mail: matsuyat@nara-edu.ac.jp.

\vskip 5.0cm
A short running head title: Singly-Peaked Population Processes

Proofs should be sent to : Toyoki Matsuyama, Department of Physics, Nara 
University of Education, Takabatake-cho, Nara 630-8528, Japan 

\end{flushleft}

\newpage

\begin{flushleft}
Abstract 
\end{flushleft}

When a small number of individuals of organism of single species is confined 
in a closed space with limited amount of indispensable resources, their 
breading may start initially under suitable conditions, and after peaking, 
the population should go extinct as the resources are exhausted. 
Starting with the logistic equation and assuming that the carrying capacity 
of the environment is a function of the amount of resources, a mathematical 
model describing such pattern of population change is obtained.  
An application of this model to typical population records, that of deer 
herds by Scheffer (1951) and O'Roke and Hamerstrome (1948), yields 
estimations of the initial amount of indispensable food and its availability 
or nutritional efficiency which were previously unspecified.  

\newpage

\begin{flushleft}
{\bf INTRODUCTION}
\end{flushleft}
\hspace{0.5cm}
   The logistic or the Lotka-Volterra model has long been a 
mathematical frame to study population dynamics tending to stationary or 
oscillating equilibrium due to intra- or interspecific interactions (e.g., 
Pielou, 1974; Begon {\it et al.}, 1996; Borrelli {\it et al.}, 1996; 
Glesson and Wilson, 1986; Reed {\it et al.}, 1996).  
Also, there is another pattern of population change which is singly-peaked.  
Typical one may be the population change of deer herds observed by Scheffer 
(1951).  
It was reported that the deer were freed in closed spaces at some definite 
time and the populations increased first nearly exponentially to reach a peak 
and then decreased or went extinct finally.  
The change was considered to be fluctuation or over-abundance from the 
sigmoidal pattern and ascribed to changes of reproduction rate and/or 
mortality due to unspecified reasons.  
But, such patterns should be generally observable if living organisms are 
confined in a closed space with constant amount of growth resources which are 
actually not reproducible although initially given.  
Effects of food availability or resource limitation on population dynamics 
are one of recent concerns 
(e.g., Ogushi and Sawada, 1985; Edgar and Aoki, 1993).
To our knowledge, however, rather few mathematical models have been studied 
to analyse such patterns of population change and the carrying capacity for 
population has been traditionally assumed to be a constant characterizing 
its environment.  
In this report, we propose a new mathematical model to interpret such a 
pattern of population change by introducing a new assumption that the 
carrying capacity is a function of the amount of resources.  
After formulation and its application to the deer herd population, we discuss 
several characters of our model as compared with the existing models.  

\begin{flushleft}
{\bf MATHEMATICAL MODEL}
\end{flushleft}
\hspace{0.5cm}
   We start with the logistic equation for a single species of organism living 
in some limited space,
\begin{eqnarray}
\frac{1}{N} \frac{dN}{dt} = r (1 - \frac{N}{K}) \ \ \ ,
\label{logistic}
\end{eqnarray}
where $N$ is the population size of the organism, $r$ the potential net 
reproduction rate and $K$ the carrying capacity of the population.  
Now we assume that the carrying capacity depends on the amount of 
indispensable resources in the space for the organisms and the resources are 
consumed by the organisms after they begin to live.  
Under such situation, we may suppose that the carrying capacity is a function 
of the amount ($X$) of the resource, $K=f(X)$.  
Then, we have
\begin{eqnarray}
\frac{1}{N} \frac{dN}{dt} = r (1- \frac{N}{f(X)}) \ \ \ .
\label{ours1}
\end{eqnarray}
We may further assume that the decreasing rate of $X$ is proportional to the 
population size and the reproduction rate of the resources is negligible 
compared with the consumption rate, i.e.,
\begin{eqnarray}
\frac{dX}{dt} = - a N
\label{ours2}
\end{eqnarray}
where $a (>0)$ is the consumption rate of the resources per individual of the 
organism per unit time. 
From Eqs.(\ref{ours1}) and (\ref{ours2}), we have
\begin{eqnarray}
\ln{\frac{N}{N_0}} = r(t-t_0) + \frac{r}{a} \int^{X}_{X_0} \frac{dX}{f(X)}
\label{integral}
\end{eqnarray}
where $N_0$, $X_0$ and $t_0$ are the initial values of $N$, $X$ and $t$, 
respectively.  
There may be various choices for $f(X)$ as an integrable function which 
represents a possible resource dependence of carrying capacity.  
We choose here the simplest one, a linear function $f(X) = bX$, with the 
proportional constant $b$($>0$), which we may call the nutritional 
efficiency.  
Then we have 
\begin{eqnarray}
N(t) = N_0 [ \frac{X(t)}{X_0} ]^{r/ab} \exp(rt) \ \ \ ,
\label{dump}
\end{eqnarray}
with $t_0=0$.  
Equation (\ref{dump}) predicts that the amount of the resources per 
individual, $X/N$, in the case of $a=r/b$, decreases exponentially with 
time from the initial value $X_0/N_0$.  
Solving the simultaneous Eqs.(\ref{ours2}) and (\ref{dump}), we obtain 
the following solutions:  
For the case $a=r/b$, 
\begin{eqnarray}
N(t) &=& N_0 \exp [ rt+ \frac{a}{r} \frac{N_0}{X_0} \{ 1-\exp(rt) \} ] \ \ \ , 
\label{solution1}
\end{eqnarray}
and for $a \not =  r/b$,
\begin{eqnarray}
N(t) &=& N_0 [1 + (\frac{a}{r} - \frac{1}{b}) \frac{N_0}{X_0} 
\{ 1- \exp(rt) \}]^{r/(ab-r)} \exp(rt) \ \ \ .
\label{solution2}
\end{eqnarray}
The $N(t)$ curve given by Eq.(\ref{solution1}) or Eq.(\ref{solution2}) has a 
single peak for a limited range or combinations of parameters $a$, $b$, 
$r$, $X_0$ and $N_0$.  
The range giving the single peak is determined from the extreme condition 
of $N(t)$.  
The solution (\ref{solution1}) in the case of $a=r/b$ has the peak if 
$rX_0/aN_0 > 1$.  
We note $rX_0/aN_0=bX_0/N_0$ in this case.  
The maximum of $N$ is given by
\begin{eqnarray}
N_m = \frac{r X_0}{a} \exp ( \frac{a N_0}{r X_0} - 1 ) \ \ \ ,
\label{max1}
\end{eqnarray}
at the time $t_m = (1/r) \ln(r X_0/a N_0)$.  
In the case of $a \not  = r/b$, the peak exists again 
when $bX_0/N_0 > 1$.  
The maximum is
\begin{eqnarray}
N_m = N_0 [\frac{1}{ab} \{ r + (ab-r) \frac{N_0}{b X_0} \} ]^{r/(ab-r)}
(\frac{r X_0}{a N_0} + 1 - \frac{r}{ab})
\label{max2}
\end{eqnarray}
with $t_m = (1/r) \ln(rX_0/aN_0 + 1 - r/ab)$.  
We show the range where the single peak exists on the $(\frac{N_0}{X_0}, b)$ 
plane in Fig. 1.  
It should be noted that our model is soluble exactly.  
We also note that it has the scale invariance under the 
change of  ($a$, $1/b$, $X_0$) into ($\lambda a$, $\lambda/b$, 
$\lambda X_0$) with an arbitrary constant $\lambda$ and the units of $X$ 
defines the units of $a$ and $b$.

\begin{flushleft}
{\bf APPLICATION TO THE DEER POPULATIONS}
\end{flushleft}
\hspace{0.5cm}
    What can be analysed by the present model?  
To show this, we apply it to the population changes of reindeer on St. Paul 
Island (SPI) from 1911 to 1950 and on St. George Island (SGI) from 1911 to 
1949 (Scheffer, 1951).  
The population data have been well-known to be of ideal observation in 
out door laboratory where the animals lived under small hunting pressure and 
were free of predator attack for the 40 years; 
the definite numbers of the animal were planted in the closed spaces at the 
definite time, after which the population showed singly-peaked changes. 
The accuracy of the numbers was estimated to be about 10 \%.  
We also apply the model to the population change of white-tailed deer at the 
George Reserve of the University of Michigan (GRM) which showed a similar 
trend from 1928 to 1947 (O'Roke and Hamerstrom, 1948).

    For the application, we need to fix one of three parameters, $a$, $b$ and 
$X_0$, and need to assume the presence of indispensable resources for the 
animal.  
We may suppose that it was lichen at least for the SPI herd.  
This is because lichen was considered to be the key forage for reindeer, 
especially in winter (Scheffer, 1951).  
The grass disappeared on SPI 40 years after the reindeer introduction, which 
was regarded as the cause of the reindeer extinction.  
We may apply Eq.(\ref{ours2}) here without adding any reproduction term for 
the plant breeding since it was reported that recovery of lichen range may 
take 15 or 20 years there.  
A caribou is reported to eat 4.5 kg of lichen a day (Bandfield, 1996).  
We infer that real values of the consumption rate of the three deer herds are 
near to this value since they belong to the same family (a Japanese deer is 
reported to eat 11 kg of grass a day).  
As the choice of the value is not so essential to obtain perspectives to 
consider the real population, we use $a=1.64$ tons a year per individual for 
the three herds commonly.  
    
    The population change $(N)$ of the SPI reindeer from Scheffer's table is 
shown in Fig. 2-A with empty circles. 
To fit the curve of Eqs. (\ref{solution2}), we use the direct search of 
optimization (DSO) for three parameters, $r$, $b$ and $X_0$ and obtain 
$r=0.182$ per year,  $b=0.111$ individual per ton and $X_0=37000$ tons. 
We notice here some deviation of the curve from the data points which might be 
caused by changes of hunting effects or weather.  
We can not clarify the reason at present, however.  
After the similar application of DSO to the population on SGI and that in GRM, 
the optimized curves are compared in Figs. 2-B and C with the observed data.  
All parameters thus obtained are summarized in TAB. I together with the areas 
of three habitats and the respective initial and maximum population sizes.  

Now we explain some characters of the population processes referring the 
figures and the table.  
The most significant result in the table is that the initial stock $X_0$ on 
SPI is more than 8 times larger than on SGI although the land areas are almost 
same.  
In the present model, the deviation of $X_0$ is proportional to that of $a$ due 
to the scale invariance for parameters mentioned above.  
However, this difference in $X_0$'s is much more than one that can be caused by 
probable difference in $a$'s.  
Rather, this may correspond to about ten times larger $N_m$ observed on SPI 
than that on SGI and suggests that SPI was much more fertile than SGI.  
Scheffer remarked some environmental differences between the two islands.  
Here we propose that the initial values of the carrying capacity is given by 
$K_0=bX_0$, of which data are also included in TAB. I.  
$K_0$ is free from the effect of $a$-ambiguity.  
The significant difference between $K_0$'s of SPI and SGI in the table also 
supports above view.  
We find next that the net reproduction rate $r$ of the SPI herd is much 
smaller than that of SGI which is further smaller than that of GRM.  
Values of $r$ are free from the effects of $a$.  
A biological reason may exist for the differences of r, although we 
cannot explain it now.  
The $b$ value of the SPI herd is about twice of that of SGI (and GRM).  
However, this difference might be caused by any difference in possible $a$ 
values.  

   Further, we find significant differences between population processes 
on SPI and on SGI (and in GRM): 
The population on SPI increased rather slowly and went extinct steeply 
after the maximum while that on SGI increased fast and decayed slowly.  
For the SPI herd, the ratio of the obtained $r$ to the $b$ value is very near 
to the $a$ value, meaning that the curve fitting for SPI reindeer is attained 
with Eq.(\ref{solution1}) or as the case of $a=r/b$, as far as the $a$ value is acceptable. 
In contrast with this, some similarities are found in the population processes 
of SGI reindeer and GRM white-tailed deer: 
The $r/b$ is much less than the assumed $a$ for both herds, meaning that the 
fitting is realized with Eq. (\ref{solution2}) or as the case of $a<r/b$.  
In spite of the large difference between the areas of two habitats, the two 
magnitudes of $b$ are nearly equal each other and the two $X_0$'s are 
too.  
Between the two habitats, a similarity in ecological characters for 
deers should have existed.  

    Now we discuss relations among observed and calculated population 
parameters.  
Inspection of the table suggests no definite relation of $r$ to $N_m$, $X_0$ 
and the respective densities.  
$r$ is presumably inversely related to $N_0$.  
The observed $N_m$ may have a linear relation to $X_0$ which is clearly found 
in Fig. 3.  
We have shown a non-linear relation between $N_m$ and $X_0$ in Eqs.
(\ref{max1}) and (\ref{max2}).  
First, for the case of  $ab/r = 1$, Eq.(\ref{max1}) is approximated 
as $N_m \approx rX_0/ea$ since $aN_0/rX_0 << 1$ within the present range of 
parameters ($e$ is the base of the natural logarithm).  
Second, for the case of $ab/r<<1$ (the SGI and GRM cases), we rewrite 
Eq.(\ref{max2}) as $N_m=N_0 (r/ab)^{r/(ab-r)}[1-(1-ab/r)N_0/bX_0]^{r/(ab-r)}$ 
$(rX_0/aN_0 + 1 - r/ab)$.  
If $ab/r+bX_0/N_0 >> 1$, we have $N_m \approx (r/a) (r/ab)^{r/(ab-r)} X_0$.  
This condition is fulfilled in the present ranges of the parameters.  
We have then a linearly-increasing trend of $N_m$ with $X_0$ for both cases.  
Concerned with the coefficients of the linear increase, we show 
the quantitative estimations of the ratio $N_m/X_0$ in TAB. II.  
$N_m({\rm DATA})$ is the maximum $N$, which was really observed on SPI, SGI, or GRM.  
$N_m({\rm LNR})$ is estimated by using $N_m \approx rX_0/ea$ for SPI case or 
$N_m \approx (r/a) (r/ab)^{r/(ab-r)} X_0$ for SGI and GRM cases.  
Eq.(\ref{max1}) or Eq.(\ref{max2}) gives us the fully theoretical value of 
$N_m$ which is denoted by $N_m({\rm DSO})$.  
We may consider that these values of coefficients are almost constant over 
three herds causing the linear relation between $N_m$ and $X_0$.  

    The minimum requirement of year-long grazing area of lichen for a reindeer 
was estimated to be 33 acres on SPI (Scheffer, 1951).  
This meant that the carrying capacity per unit area was 0.030 and the carrying 
capacity of SPI was 800 individuals (Dasman, 1964).  
The peak densities ($N_m$/area, estimable in TAB. I) exceeds 0.03 in two 
habitats, SPI and GRM, which were considered to be fluctuations over the 
carrying capacity.  
In our model, we postulate that the carrying capacity is not a constant of a 
land but a changeable parameter which depends on environmental conditions, 
e.g., the quantity of indispensable forage for the animal.  
Referring the $K_0$, the initial value of the carrying capacity defined above, 
 we find $N_m \leq K_0$ in TAB. I, a reasonable limiting relation of the maximum 
population to the maximum carrying capacity.

    Finally, we compare the present model with the original Lotka-Volterra 
system (LVS) for predator -prey interaction.  
In fact, at a glance, the deer may be regarded as predator and the lichen as 
prey.  
The system is given by 
\begin{eqnarray}
\frac{dP}{dt}=-cP + \alpha PS \ \ \ ,
\label{predator}
\end{eqnarray}
for the predator population size ($P$) and 
\begin{eqnarray}
\frac{dS}{dt}=kP - \beta PS \ \ \ ,
\label{prey}
\end{eqnarray}
for the prey population ($S$) with the coefficients $c$, $k$, $\alpha$ and 
$\beta$ of the well-known meanings (Borrelli {\it et al.}, 1996).  
When $k=0$, this system becomes that of the ordinary differential equations of 
Kermack-Mckendrick type (KMS) and can reproduce a singly-peaked process if the 
initial value of $S$ is larger than $c/\alpha$.  
However the LVS or KMS contains as its essence the encounter term which is 
proportional to $PS$.  
This means that encounter between two interacting species should take place 
with a constant probability uniformly through-out the space and time 
(applicability of the mass-action law).  
Hence the system should be applicable to the case of thin populations of prey 
and predator.  
Our model has no such encounter term (see Eq. (\ref{ours2})) to lead such limitation 
to the population density.  
The estimated values of $X_0$ or $X$ per unit area may be interpreted to be of 
thin or dense population (or stock) of the lichen (or forage) according to its 
magnitude.  
For the predator or deer, the present model assumes only intraspecific 
competition as the original logistic does.  
Hence the estimated values of $N_m$ and $r$ may be of dense population.  
Of course, effects of overcrowding can be discussed within LVS by 
introducing the $S^2$ and $P^2$ terms to it.  
However, an addition of new terms with new parameters may make the analysis 
more vague unless the parameters are determined by any other methods.  
We should also note that the unimodal curves can be reproduced by a modified 
logistic equation with a term of integrated toxins for population 
(Small 1987).  
However, the model has no explicit relationship with the resources for the 
population.  

\begin{flushleft}
{\bf CONCLUSIONS}
\end{flushleft}
\hspace{0.5cm}
    We have presented a simple mathematical model with which one can analyse 
singly-peaked population processes.  
Although it is simple, the model provides a good account of the deer 
population dynamics by assuming the resource-dependent carrying capacity and by 
introducing two ecological parameters, the consumption rate of 
indispensable resources ($a$) and the nutritional efficiency ($b$), in 
addition to such traditional ones as the reproduction rate ($r$) and the 
initial stock of the indispensable resources $X_0$.  
Here $a$ and $b$ can be in principle determined by observation.  
The model is soluble exactly as the original logistic is, providing 
mathematical benefits.  
It may be applicable to consumption of fertilizer by plant (perfectly zero 
breeding of prey) and to the case of non-zero breading of prey by adding a 
breding term for it in Eq. (\ref{ours2}).  
Further we add that a population can go extinct steeply, or even suddenly, 
from its peak in the model.  
Breeding and extinction of many organisms should depend or should have 
depended on their indispensable resources of finite amount to which processes 
the present model may be applied. 

\begin{flushleft}
{\bf ACKNOWLEDGMENTS}
\end{flushleft}
\hspace{0.5cm}

We would like to thank Prof. H. Sato (Dept. of Biology, Nara Women's Univ.) 
and Prof. N. Kitagawa (Dept. of Biology, Nara Univ.of Education) for 
introducing us some literatures and informations on related subjects.  
We also sincerely thank to the referee for giving us very useful comments.  

\newpage
\begin{flushleft}
{\bf REFERENCES}

Banfield, A. W. F. 1996. Caribou in the "Encyclopedia americana", Grolier 
Incorp., 
\hspace*{0.3cm} Danbury, Connecticut, p.659.          

Begon, M., Mortimer, M. and Thompson, D. J. 1996. "Population Ecology", 
Blackwell \hspace*{0.3cm} Science, Oxford.

Borrelli, R. L. and Coleman, C. S. 1996. "Differential Equations, a modeling 
\hspace*{0.3cm} perspective", John Wiley \& Sons, NY. 

Dasmann, R. F. 1964. "Wildlife Biology," John Wiley \& Sons, New York, NY.

Edgar, G. J., and Aoki, M. 1993. Resource limitation and fish predation: their 
\hspace*{0.3cm} importance to mobile epifauna associated with Japanese 
Sargassum, {\it Oecologia} {\bf 95}, 
\hspace*{0.3cm} 122-133.

Gleeson, S. K. and Wilson, D. S. 1986. Equilibrium diet: optimal foraging and 
prey \hspace*{0.3cm} coexistence, {\it Oikos} {\bf 46}, 139-144.

Ogushi, T. and Sawada, H. 1985. Population equilibrium with respect to 
available food 
\hspace*{0.3cm} resource and its behavioural basis in an herbivorous lady 
beetle, henosepilachna 
\hspace*{0.3cm} niponica, {\it J. Anim. Ecol.} {\bf 54}, 781-796.

O'Roke, E. C. and Hamerstrom, Jr., F. N. 1948. Productivity and yield of the 
George 
\hspace*{0.3cm} reserve deer herd, {\it J. Wildl. Mgmt.} {\bf 12}, 78-86.

Pielou, E. C. 1974. "Population and community ecology," Gordon and Breach 
Sci.Pub., \hspace*{0.3cm} New York, NY.

Reed, D. J., Begon, M. and Thompson, D. J. 1996. Differential cannibalism and 
\hspace*{0.3cm} population dynamics in a host-parasitoid system, {\it Oecologia} {\bf 105}, 189-193.

Scheffer, V. B. 1951. The rise and fall of a reindeer herd, {\it Scientific 
Monthly} {\bf 73}, 356-362.

Small, R.D. 1987.  Population growth in a closed system, in "SIAM Mathematical 
\hspace*{0.3cm} Modelling: Classroom Notes in Applied Mathematics (ed. by M.S. Klamkin)", SIAM, \hspace*{0.3cm} 317-320, Philadelphia.

\end{flushleft}

\newpage

TAB. I.  Population data of three deer herds: The habitat area, the initial 
and maximum population sizes ($N_0$  and $N_m$) are from references (Scheffer, 
1951; O'Roke and Hamerstrom, 1948).  
The nutritional efficiency ($b$), the initial stock of indispensable food 
($X_0$) and the reproduction rate ($r$) which are defined in text are obtained 
in this work after direct search optimization of the theoretical curve 
(Eqs.(\ref{solution1}) and (\ref{solution2}) in text) to fit the population 
records in the references and shown with significant figures of three digits.  
The consumption rate of indispensable food ($a$) is fixed to be 1.64 tons per 
year per individual for three herds.  
$K_0 (=bX_0)$ is the initial carrying capacity given in text.  

\vskip 1.5cm

\noindent
\begin{tabular}{lccccccc} 
\hline
Herd&Habitat area&$N_0$&$N_m$&$r$       &b           &$X_0$&$K_0$
\\
    &(acre)      &     &     &($y^{-1}$)&(indiv./ton)&(ton)&
\\ \hline
St.Paul I. reindeer&26500&25&2046&0.182&0.111 \ &37000&4090
\\ 
St.George I. reindeer&22400&15&\ 222&0.469&0.0512&\ 4460&\ 229
\\ 
George Res. w.t.deer&\ 1200&\ 6&\ 211&0.740&0.0561&\ 4150&\ 233
\\ \hline
\end{tabular}
\vskip 1cm
\centerline{TAB. I.}  

\vskip 1.5cm

TAB. II.  The estimations of $N_m/X_0$ of three deer herds: 
$N_m({\rm DATA})$, $N_m({\rm LNR})$, and $N_m({\rm DSO})$ are defined in 
text.  
They are divided by $X_0$ which takes the value corresponding to each herd 
in TAB. I.

\vskip 2cm

\begin{tabular}{lccc} 
\hline
Herd&$N_m({\rm DATA})/X_0$&$N_m({\rm LNR})/X_0$&$N_m({\rm DSO})/X_0$
\\ \hline
St.Paul I. reindeer&0.0553&0.0407&0.0409
\\ 
St.George I. reindeer&0.0498&0.0352&0.0356
\\ 
George Res. w.t.deer&0.0508&0.0417&0.0419
\\ \hline
\end{tabular}
\vskip 1cm
\centerline{TAB. II.}

\newpage
\begin{flushleft}
{\bf Figure Captions}

\vskip 1cm

FIG. 1. The range where the condition giving a peak in $N(t)$ curves is 
fulfilled:  $b>N_0/X_0$.  Notations are defined in text.  

\vskip 1cm

FIG. 2. Population curves obtained with Eqs.(\ref{solution1}) and 
(\ref{solution2}) in text to fit the population records of the three deer 
herds.  The nutritional efficiency (b) and the initial stock of indispensable 
food ($X_0$) and the potential reproduction rate ($r$) which are defined in 
text are optimized. 
Plate A, reindeer on St. Paul Island (Scheffer, 1951); 
Plate B, reindeer on St. George Island (ibid); 
Plate C, white-tailed deer in George Reserve Michigan (O'Roke and Hamerstrom, 
1948).

\vskip 1cm

FIG. 3.  The maximum size of deer populations observed ($N_m$) vs the 
initial amount of indispensable food ($X_0$) estimated in text.  
\end{flushleft}

\newpage
\begin{figure}
\epsfysize=7cm
\centerline{\epsfbox{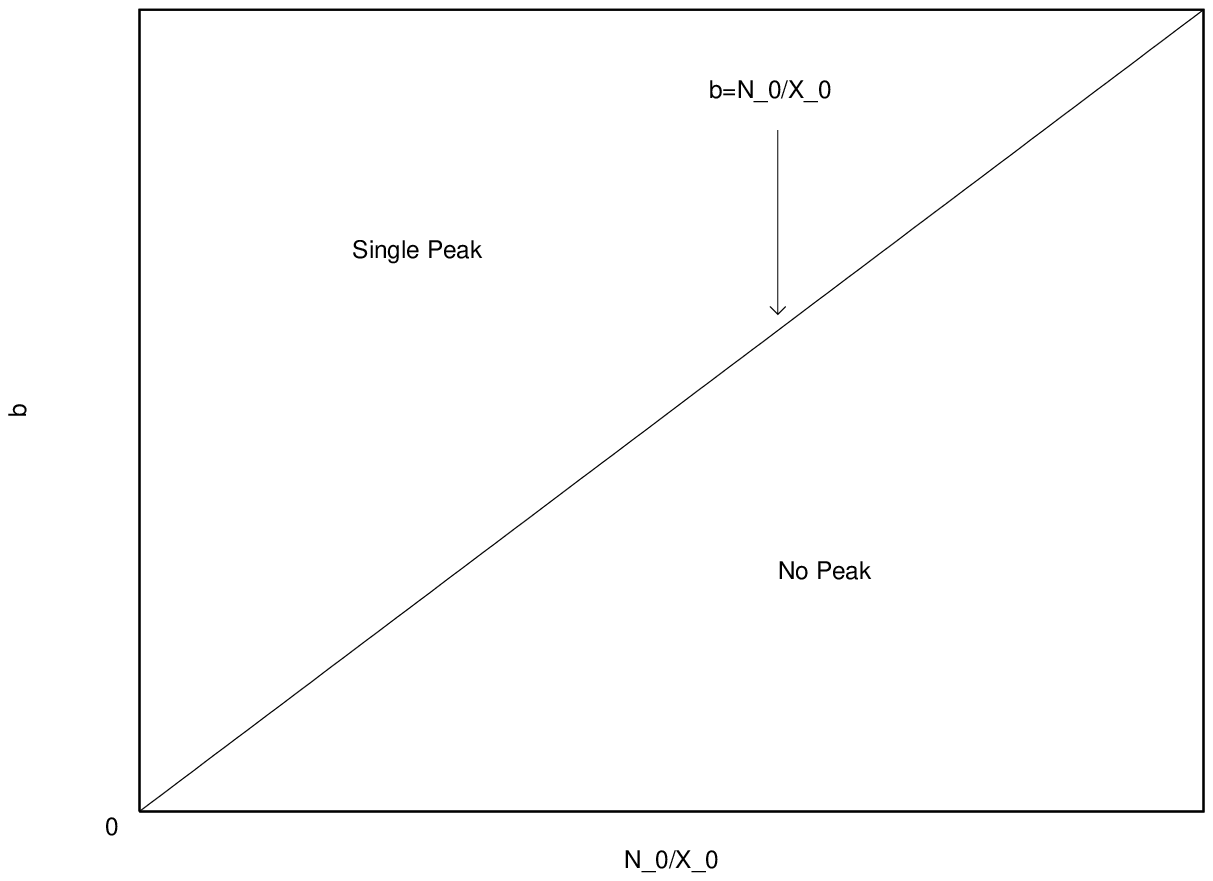}}
\centerline{FIG. 1}
\end{figure}

\newpage
\begin{figure}
\epsfysize=7cm
\centerline{\epsfbox{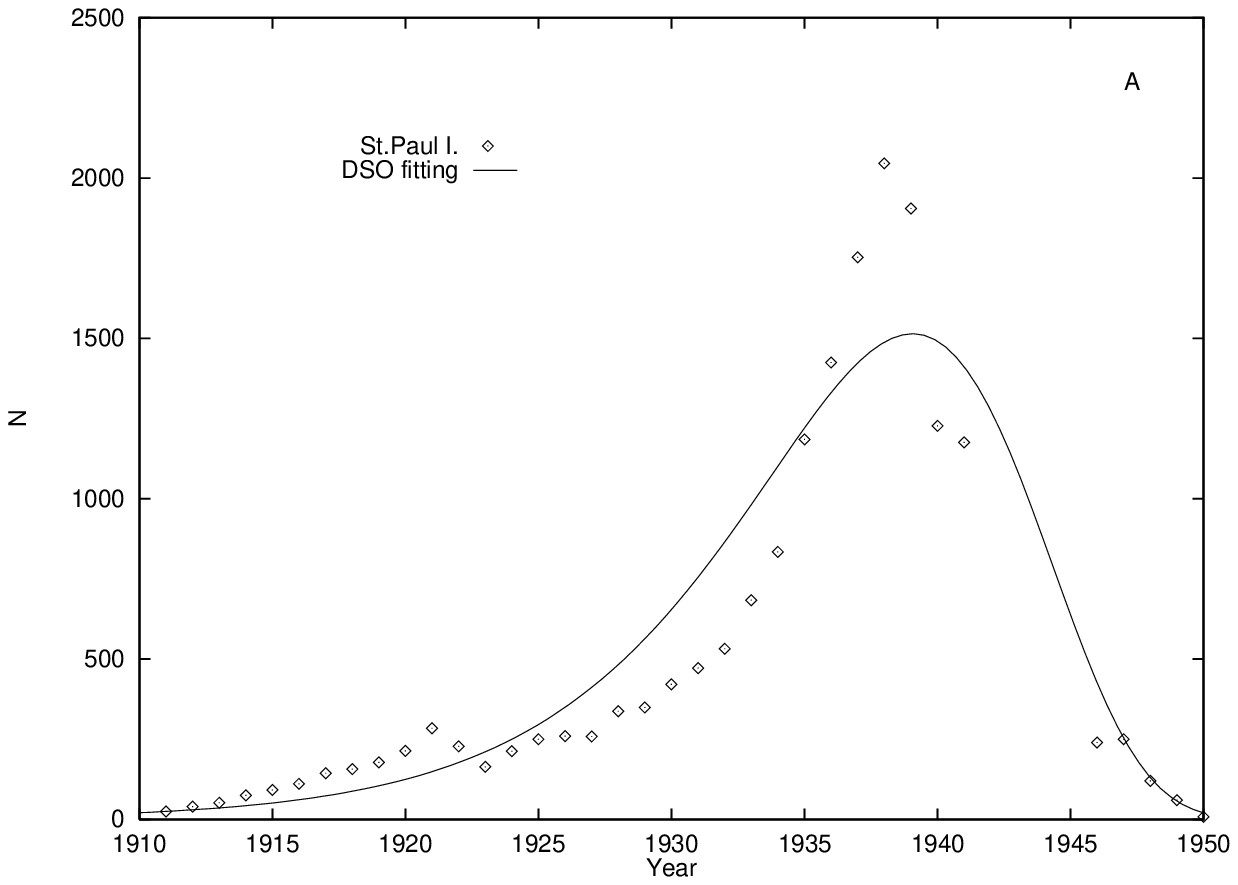}}
\centerline{FIG. 2A}
\end{figure}

\newpage
\begin{figure}
\epsfysize=7cm
\centerline{\epsfbox{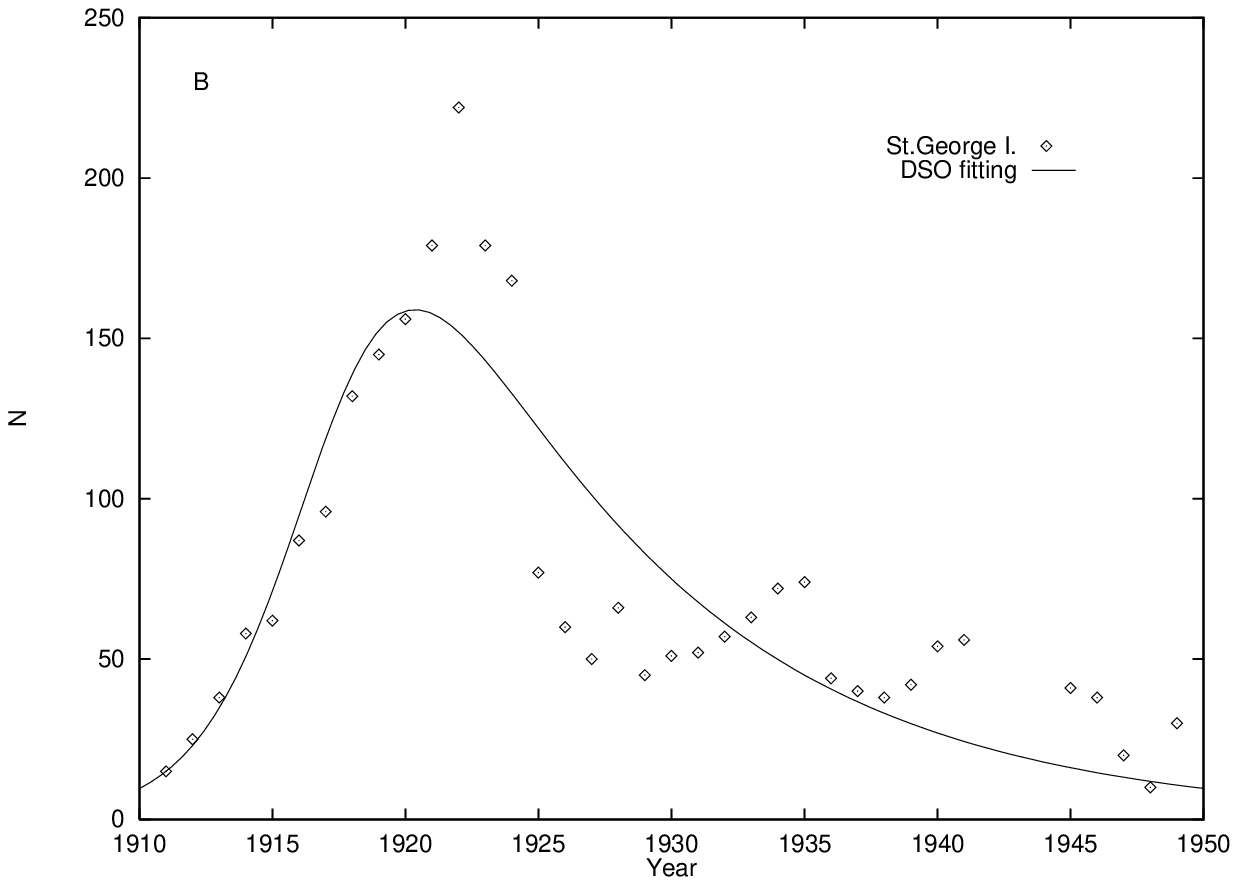}}
\centerline{FIG. 2B}
\end{figure}

\newpage
\begin{figure}
\epsfysize=7cm
\centerline{\epsfbox{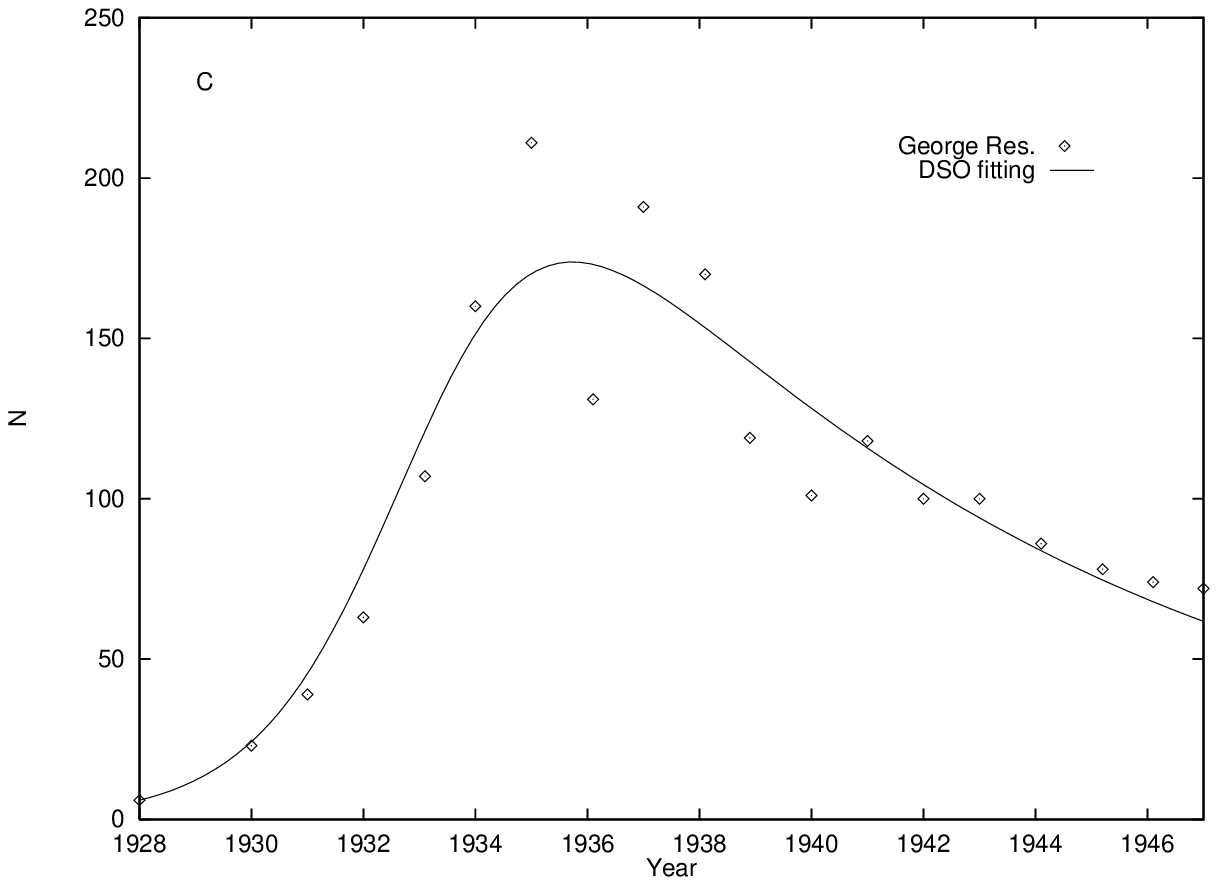}}
\centerline{FIG. 2C}
\end{figure}

\newpage
\begin{figure}
\epsfysize=7cm
\centerline{\epsfbox{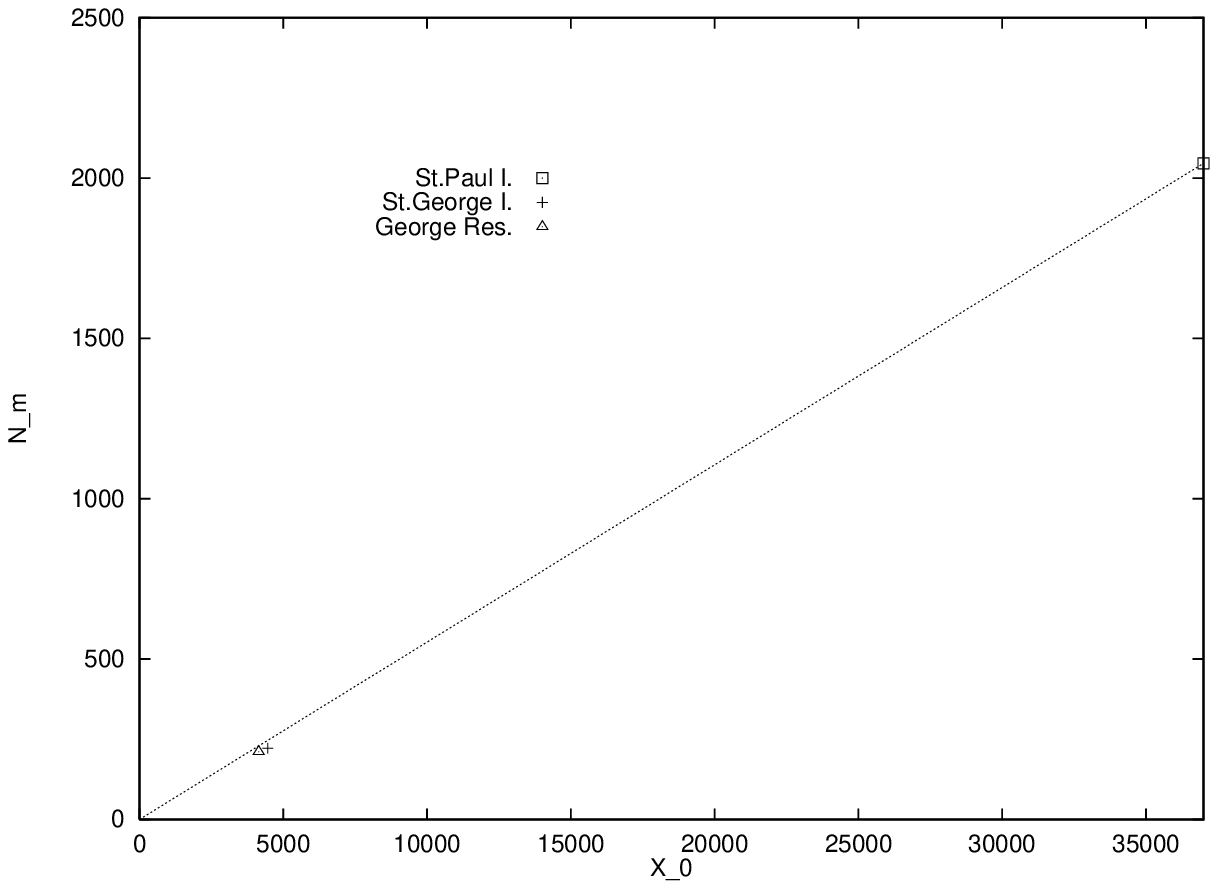}}
\centerline{FIG. 3}
\end{figure}

\end{document}